\documentclass[preprint]{aastex}
\usepackage{natbib}
\usepackage{graphicx}
\usepackage{subfigure}
\usepackage{amsmath,amssymb}

\begin{document}

\title{Heating of Flare Loops With Observationally Constrained Heating Functions}
%
%
%
\author{Jiong Qiu$^{1}$, Wen-Juan Liu$^{1}$, Dana W. Longcope$^{1}$}
\affil{1. Department of Physics, Montana State University, Bozeman MT 59717-3840}

\begin{abstract}
We analyze high cadence high resolution observations of a C3.2 flare obtained by 
AIA/SDO on August 1, 2010. The flare is a long duration event with soft X-ray and EUV
radiation lasting for over four hours. Analysis suggests that magnetic reconnection and 
formation of new loops continue for more than two hours. 
Furthermore, the UV 1600\AA\ observations show that each of the individual 
pixels at the feet of flare loops is brightened instantaneously 
with a timescale of a few minutes, and decays over a much longer timescale of more 
than 30 minutes. We use these spatially resolved UV light curves during the rise phase to construct
empirical heating functions for individual flare loops, and model heating of coronal plasmas 
in these loops. The total coronal radiation of these flare loops are compared with soft X-ray 
and EUV radiation fluxes measured by GOES and AIA. This study presents a method to observationally 
infer heating functions in numerous flare loops that are formed and heated 
sequentially by reconnection throughout the flare, and provides a very useful constraint to coronal heating models. 
\end{abstract}

\section{INTRODUCTION}
Magnetic reconnection is the commonly accepted mechanism governing energy release in solar
flares. Reconnection allows a change of connectivity between magnetic structures
and therefore the overall topology and magnetic energy in the field. 
To achieve magnetic energy release it is necessary that the post-reconnection magnetic 
field contains less energy than the pre-reconnection configuration. The variety of radiation 
signatures observed in flares demands that a significant amount of excess magnetic energy 
be released to heat plasmas, directly or indirectly, in post-flare loops.  It has been known 
for decades that flare radiation in $> 20$ keV hard X-rays, commonly recognized as a product 
of non-thermal electron beams colliding with the lower-atmosphere (so-called thick target), 
usually exhibit an impulsive time history. This emission has typically ended by the time 
soft X-rays (such as observed by GOES) reach their maximum. This is the well known Neupert 
effect \citep{Neupert68}, the physical basis of which is that non-thermal beams deposit most, 
if not all, energy, in the lower atmosphere, driving chromospheric
evaporation to fill as well heat post-flare loops. Observations in the EUV bands further
reveal gradual time profiles that are delayed with respect to soft X-ray time profiles.
The decay phase of the soft X-ray radiation and the delayed radiation in the EUV bands (typically
sensitive to up to a few million K degrees) are considered to be caused by cooling of hot plasmas
which have been heated during the impulsive phase. 

Theoretic models have been developed to investigate hydrodynamic evolution of plasmas inside the flare
loop, from which flare radiation at different wavelengths can be calculated and compared with
observations by satellites. It is evident that how the flare loop is heated, namely, when, where,
for how long, by how much, and in what form, governs hydrodynamic evolution of plasmas inside the flare loop.
Discovery of the Neupert effect has motivated investigation of heating flare plasmas with energy flux
that is released during the impulsive phase, for example, as carried by non-thermal beams
\citep[e.g.,][]{Fisher90}. However, a couple of discrepancies have since then arisen. First,
it has been found that the observed long decay in soft X-ray and EUV
radiation cannot be explained by a pure cooling process, which is either conductive or radiative
cooling as traditionally considered. Recently, \citet{Bradshaw10a, Bradshaw10b} 
have modeled corona cooling by enthalpy flow during the decay phase of the flare, which transfers
energy (and mass) to the transition region to be radiated away there. Cooling by the 
enthalpy flow becomes dominant in the decay phase in long tenuous loops where radiative 
cooling is less efficient. 

Along another avenue, it has become recognized that energy 
release may continue well into the traditionally called decay phase of the flare, when 
impulsive hard X-ray emission has ended \citep[e.g.][]{Bornmann85}. \citet{Cargill83} proposed that a post-flare loop 
may be heated by MHD waves for an extended time. The idea of continuous heating during the decay
phase is further supported by the second kind of observational discoveries. \citet{Czaykowska99, Czaykowska01}
found blue-shift signatures at the outer edge of ``moving" flare ribbons in the decay phase, 
indicative of chromospheric evaporation responding to energy release in newly formed flare 
loops. These observations provide evidence of continuous energy release by magnetic reconnection 
in the decay phase of the flare, and the energy transfer that drives evaporation is by thermal conduction. 
Recent high resolution EUV imaging observations \citep[e.g. ][]{Aschwanden01} by the Transition 
Region And Corona Explorer \citep[TRACE;][]{Handy99} 
have indeed shown numerous post-flare loops with cross-sectional area of $\le$ 1\arcsec. It is further 
confirmed by \citet{Fletcher04} that the flare UV ribbons are made of small kernels outlining the feet
of flare loops. These loops are formed by so-called ``patchy" reconnection \citep{Longcope09}
and heated at different times during the flare, and evolve
independently of each other. The observed unresolved coronal radiation flux at any given time
is the sum of radiation by these loops at different evolution stages with different temperatures
and densities. To correctly model heating of flare plasmas, the challenge has been to
identify the temporally and spatially resolved energy release process in individually formed
and heated flare loops, and this process may proceed well into the decay phase of the flare.

In view of these recent observational discoveries, \citet{Longcope10} have developed 
a model to compute energy release rate in the 3-dimensional patchy reconnection events 
along a current sheet and implemented it in analyzing a two-ribbon flare observed by TRACE. 
The patchy reconnection forms many flare loops that are identified in TRACE high resolution imaging
observations at EUV 171~\AA. Each of these loops is heated by
an amount of energy calculated from the reconnection model using reconnection flux
measured from observations of flare ribbon evolution. This
comprehensive study combining modeling and observations in a few novel ways
has successfully reproduced the synthetic coronal radiation fluxes that agree with those
observed by GOES and RHESSI, and paved a new avenue to relate reconnection physics with
plasma heating in flares.

In the present study we introduce and experiment with an alternative method to observationally 
infer heating rates in flaring loops by analyzing spatially resolved UV emission time profiles.
Analysis of joint hard X-ray and UV observations by SMM has revealed a close temporal correlation
between hard X-ray and UV light curves in solar flares \citep{Cheng81, Cheng88}. 
On the other hand, whereas high resolution UV 1600\AA\ observations by TRACE
confirm the strong temporal correlation between the UV and hard X-ray counts,
it does not always show spatial coincidence between the two \citep{Warren01, 
Coyner09, Cheng12}. These observations suggest the possibility of heating the 
chromosphere by either electron beams or thermal conduction, both producing the rapid 
rise of UV emission though likely by different amounts \citep{Cheng12}.
It therefore appears that the spatially resolved UV light curve provides information of temporally resolved
energy release rate in newly formed flare loops. This method, if successful, will 
enable us to examine reconnection energy release from both modeling and observational points of view. In this paper, we apply
this method to a C3.2 flare observed by the Atmosphere Imaging Assembly \citep[AIA;][]{Lemen11}
onboard the {\em Solar Dynamic Observatory} (SDO). In the following Section 2, we describe the observations
of the flare. In Section 3, we construct temporally resolved heating rates, which are used
to model plasma evolution in flare loops, and compare computed synthetic coronal radiation
with observations. In Section 4, we study how model parameters affect the predicted 
coronal radiation light curves. Conclusions and discussions are given in the last section.

\section{OBSERVATIONS AND ANALYSIS}
\subsection{Overview of Observations}
\begin{figure}
\epsscale{0.7}
\plotone{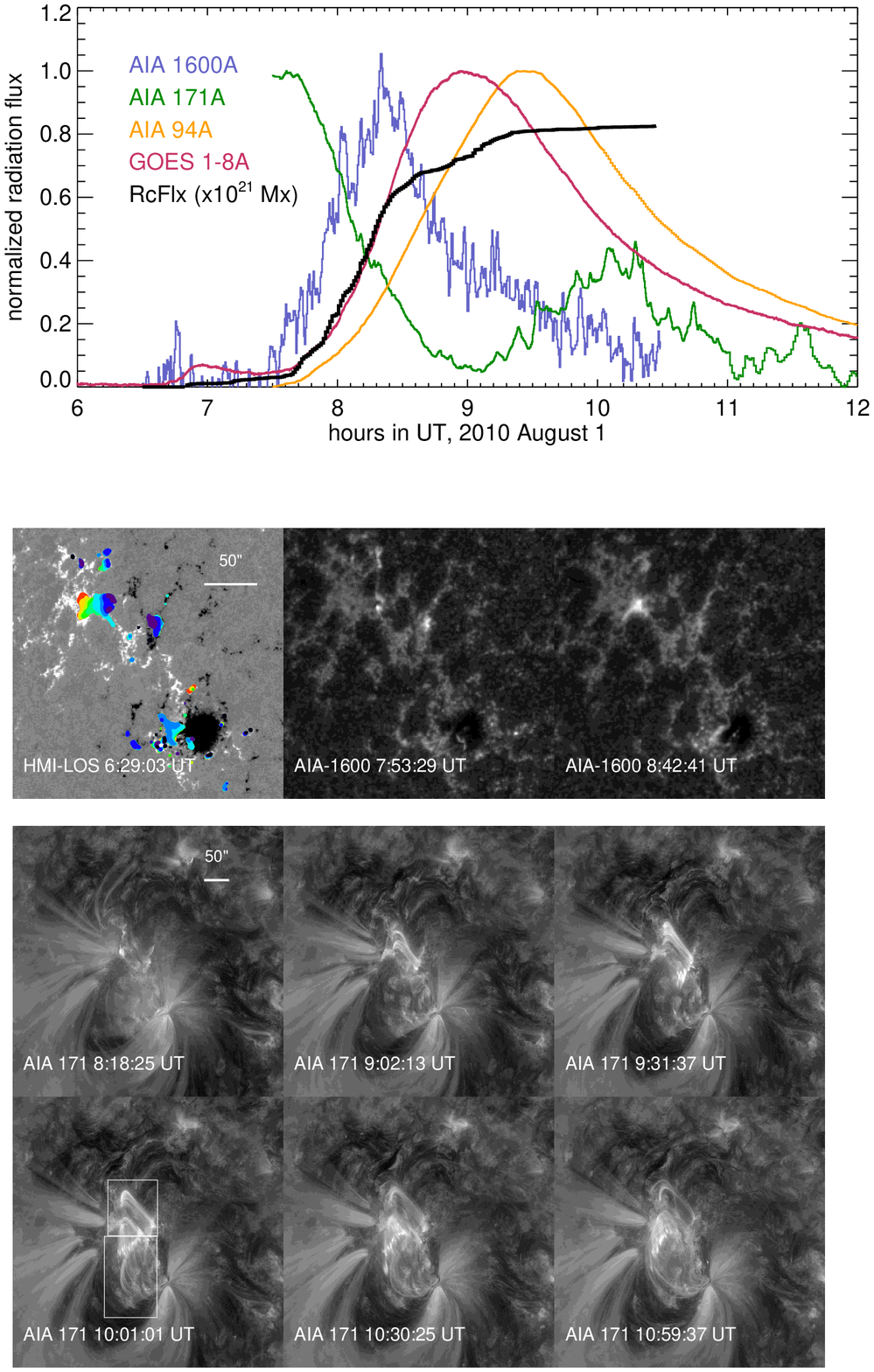}
\caption{Upper: light curves of the 2010 August 1 C3.2 flare in UV 1600\AA\ and EUV 171\AA\ and 94\AA\ by SDO/AIA, 
and soft X-ray 1-8\AA\ by GOES. Also plotted is the reconnection flux measured using UV 1600\AA\ images observed
by AIA and the longitudinal magnetogram by HMI. Middle left: evolution of the UV brightening in the lower atmosphere.
The colors from purple to red indicate times from 7:30 to 9:30 UT. Middle right: snapshots of UV images at two
different times observed by AIA showing brightening at the foot-points of the first and second sets
of flare loops respectively. Lower: snapshots of EUV 171\AA\ images at six different times showing
the impulsive phase, the first set of shorter post-flare loops, the second set of longer post-flare loops, 
and their evolution. The two boxes in the lower left panel indicate two regions where the EUV fluxes are summed. 
} \label{Fig1}
\end{figure}

The focus of this work is the C3.2 flare on 2010 August 1 summarized in Figure~\ref{Fig1}.  Colored curves in the 
top panel show normalized light curves from several bands: broad-band soft X-ray (SXR) from GOES, narrow EUV bands 
at 94\AA\ and 171\AA\ from SDO/AIA, and broad-band UV at 1600\AA\ also from SDO/AIA.  Each of these light curves 
peaks at a different time; the earliest is the UV curve which begins rising at 7:30UT and peaks at 8:18UT.  
This is understood in terms of a flare model in which magnetic energy is released in the corona, and deposited into 
the chromosphere enhancing UV emission and evaporating hot plasma into the post-reconnection coronal loops.  
The coronal plasma then cools down, becoming visible sequentially as loops in the SXR and EUV bands 
characteristic of successively diminishing temperatures. The total UV count rate light curve in Figure~\ref{Fig1} 
exhibits fluctuations on short timescales, which might suggest heating by electron beams; previous studies 
have revealed temporal correlation between hard X-ray and UV light curves \citep{Cheng88}. However, 
RHESSI \citep[][; data not shown here]{Lin02} detects little evidence of emission above 20~keV, and the X-ray 
light curves by RHESSI evolve rather gradually, so it appears that most of the initial energy transport was from 
thermal conduction. It is very likely that heating by non-thermal particles is present, but the
non-thermal flux in this small C3.2 flare may be lower than the detection threshold of RHESSI.
We, therefore, take the approximation that heating by precipitating, non-thermal electrons is 
insignificant in this particular case, and neglect it in our modeling. 

The UV emission from the lower-atmosphere holds the key to a deeper understanding of this scenario since it is the most direct evidence of the energy release initiating it.  Its spatial structure provides information about the energy release geometry.  The sequence of AIA 1600\AA\ images, with 
0.6\arcsec\ pixel scale and a 24~s cadence (see $e.g.$ middle-right panel in Figure \ref{Fig1}), are coaligned with one another.  The sequence is then coaligned with a single magnetogram by Helioseismic and Magnetic Imager \citep[HMI;][]{Schou2011} onboard SDO, from 30 minutes prior to flare onset, after correcting for solar rotation. The AIA images are then rebinned to pixel scale of 1\arcsec\ by 1\arcsec.   The middle left panel of Figure \ref{Fig1} shows that brightened UV pixels forming the flare ribbons spread outward, outlining feet of numerous flare loops formed progressively at growing altitudes.  The magnetic flux swept up by the ribbons may be integrated in the HMI magnetogram to yield a plot of reconnected flux versus time \citep{Qiu02, Longcope07} plotted as a black curve in the top panel.  This rises to 8$\times 10^{20}$ Mx over 2 hours.

Both the reconnection flux and the accompanying light curves make clear that this small C3.2 flare is a long-duration event with emission timescales ranging from 2 to 4 hours; reconnection continues to add new flux for over 2 hours.  The slowly evolving effects are, however, composed of numerous small, rapid energy releases which occur as individual loops are formed by reconnection.  The coronal emission characterizing the later phases of the flare consists of thin loops (see bottom panels of Figure \ref{Fig1}).  These loops fall into two categories suggesting a two-phase flare.  The initial phase consists of shorter loops (mean foot-point separation of
about 50\arcsec) in the North.  The second phase, beginning around 9:30~UT,  consists of longer loops (foot-point separation of over 100\arcsec) in the South.  The negative footpoints of these loops fall in distinct flux concentrations shown in the middle left panel of Figure \ref{Fig1}. 

\subsection{Charactertistics of UV Light Curves}
\begin{figure}
\epsscale{1.0}
\plotone{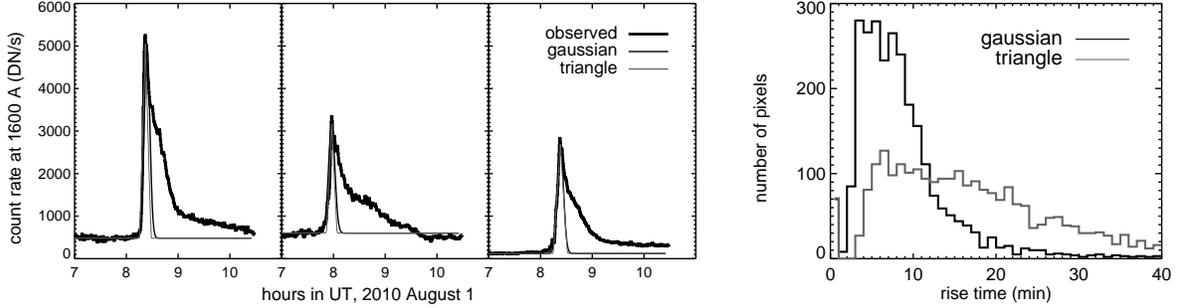}
\caption{Left: observed UV 1600\AA\ count rate light curves (thick dark) of three different flaring pixels.
The impulsive rise phase of each light curve is fitted to a gaussian or triangle function (grey curves), which
are then used to construct heating functions (see text). Right: histogram of rise time of UV light curves in 
2484 pixels from the fit to Gaussian (FWHM) or triangle functions (see text).}\label{Fig2}
\end{figure}

The logic of the above flare scenario suggests that the
light curves of individual UV pixels reflect the amplitudes, timing and durations of individual energy release events.  Three examples in Figure \ref{Fig2} show rapid rise followed by a much slower decay lasting an hour or more characteristic of UV brightening in a single pixel (after subtracting pre-flare intensity).  The rapid rise reflects the deposition of energy into the lower atmosphere, while the slow decay is a manifestation of subsequent cooling.  We thus propose that the rise time, $\tau$, and peak amplitude $I_0$, of the UV pixel light-curve are indications of the rate of energy deposition into the loop anchored to that footpoint.

The behavior of these three pixels is typical of all flare-driven UV brightening.
Previous studies, using TRACE observations at 1600\AA, revealed that the majority of flaring pixels are brightened instantaneously with a typical rise time of a few minutes, and then undergo a longer decay \citep{Qiu10, Cheng12}. This is also observed for the C3.2 flare in this study as shown in Figure~\ref{Fig2}. 

The rise-phase count rate $I(t)$ (in units of DN s$^{-1}$), after subtracting pre-flare emission, can be fitted to a half Gaussian
$$I(t) = I_0 {\rm exp}[\frac{-(t - t_0)^2}{2\tau^2}],  (t < t_0) $$ 
where $I_0$ is the peak count
rate, $t_0$ the peak time, and $\tau$ the characteristic rise time.  Alternatively, the same light curve could be fitted to a linear function 
$$I(t) = I_0(1 + \frac{t - t_0}{\tau}),  (t_0 - \tau < t < t_0) $$ 
where $\tau$ is the rise
time from the pre-flare to the peak. Fits of both kinds to the observed light curve are shown
in Figure~\ref{Fig2} by grey lines.  The panel on the right shows histograms of the rise times
from fits to all the flaring pixels.  This shows that the majority of flaring pixels rise over a time less than  10 minutes regardless of which method is used to quantify it.

Since the lower atmosphere (transition region and chromosphere) responds to 
impulsive energy deposit very efficiently on timescales of less than 1~s 
\citep{Fisher85, Canfield87}, we consider that the rise of the UV emission reflects a single energy release episode lasting for a few minutes in the flux tube (or a bunch of finer flux tubes) rooted 
at the pixel. On the other hand, the long decay
of the UV emission in the 1600\AA\ is most likely the contribution of C {\sc iv}, a
transition region line formed at 10$^5$ K degrees. Previous observations and radiative transfer modeling reveal that the C {\sc iv} emission during the decay phase reflects the evolution of overlying coronal plasma;  C {\sc iv} is therefore referred to as the coronal pressure gauge \citep{Hawley92, Griffiths98, Hawley03}.

Analysis of the UV light curves in individual pixels therefore provides quantitative information about the energy release and the subsequent plasma evolution in each coronal flux tube. In particular, we take the start-time of the UV brightening as the onset of the reconnection event forming the new flux tube. The rise time of the UV brightness gives the duration of the impulsive energy release in the newly formed tube, and the maximum brightness of the pixel reflects the magnitude of the energy release (or heating rate) in the flux tube; this simply assumes that a brighter pixel is more strongly heated. These observationally measured quantities may be then used to construct heating rates and study subsequent plasma evolution inside flare loops that 
are formed and heated sequentially during the flare. We propose this as a a preliminary, expedient 
approach to the problem, to be tested in the present experiment. Further discussion of the method will 
be presented in the last section.

The generally-accepted model holds that the time evolution of the coronal loop following reconnection energy release, 
occurs through evaporation and cooling by thermal conduction, radiation, and enthalpy flux (in the decay phase).  
If this model is correct then the radiative signatures from coronal plasma, SXR and EUV emission, can 
be predicted directly using the UV emission to infer the initial heating event.  We follow this procedure below, solving time-dependent equations for the time-evolution of the coronal plasma following an impulsive heating event.

\section{MODELING PLASMA EVOLUTION IN FLARE LOOPS WITH DISCRETE HEATING FUNCTIONS}
\subsection{Loop Evolution Via EBTEL}

We use the zero-dimensional, time-dependent model called the enthalpy-based thermal evolution loop model \cite[EBTEL;][]{Klimchuk08} to solve for the coronal response to a specified energy input.  This is a generalization of models by \citet{Antiochos1978} and \citet{Cargill1995} where characteristic density and pressure variables change in response to heating and cooling.  The loop geometry is fixed and assumed small enough that gravity can be neglected.
In this study, we construct the empirical heating function at each of the few thousand pixels, regardless of connectivity. 
In other words, in our model, we are heating numerous half loops anchored at positive or negative magnetic fields. Therefore, only the half-length of the loop, $L$, is pertinent.   \citet{Klimchuk08} define the loop to end, and the transition region to begin, at the point where downward thermal conductive flux peaks.  Above this point, in the loop, thermal conduction is an energy loss, while below it is an energy source.  Overall it does not change the energy, which is represented using pressure.

The novel element of the EBTEL model is to couple thermal conduction to evaporation through enthalpy flux across the ends of the loop (from the transition region).  The energy loss (for example, by radiation) through the transition region itself is assumed to be more effective, by a specified factor $c_1$, than coronal radiation.  If this loss is not exactly balanced by the conductive input from the corona then the difference, either positive or negative, is advected into the corona as an enthalpy flux.  This defines a velocity which also carries mass either upward, as evaporation, or downward, as condensation.  In this manner, excess heating in the corona leads indirectly, through enthalpy flux, to evaporation which then boosts the direct radiative cooling of the corona.

The EBTEL model is driven by a time-dependent volumetric heat input, $Q(t)$, into the corona.  Since the model lacks spatial resolution there is no need to identify the location at which this heat is deposited, provided it is added to the corona.  The model allows separately for heat to be added directly to the transition region via a flux of non-thermal electrons.  This will create an immediate imbalance driving evaporation more directly than will heat added to the corona, which drives evaporation through thermal conduction.  

The C3.2 flare currently under consideration does not exhibit impulsive X-ray emission at high energies (e.g., $\ge$ 20 keV), so we
consider that the flare did not produce significant amount of non-thermal electrons, and neglect the non-thermal energy input.  
Electron number density and pressure therefore evolve according to the EBTEL equations
\begin{eqnarray}
  {dn\over dt} &=& -{2c_2\over 5k_{\rm B}T}\, \left[ {F_c\over L} + c_1n^2\Lambda(T) \right] 
  \label{EBTEL_eq1} \\
  {dp\over dt} &=& {2\over 3}\,\left[ Q(t) - (1+c_1)n^2\Lambda(T) \right]
  \label{EBTEL_eq2}
\end{eqnarray}
where $c_2=0.87$ is the typical ratio of mean temperature, $T=p/2k_{\rm B}n$, to peak temperature, $\Lambda(T)$ is the radiative loss function, and $F_c$ is the conductive flux.
Thermal conduction is computed with a classical Spitzer-Harm form,
$$ F_c = -\hbox{${2\over 7}$}\,\kappa_0 (T/c_2)^{7/2}/L$$
but can be capped at a saturated value.  It enters the density evolution rather than pressure evolution since it does not remove thermal energy but rather transports it from coronal to transition region to drive enthalpy flux and with it mass flux.

The EBTEL model has been benchmarked against more sophisticated gas-dynamic models with spatial resolution along the loop.  The time evolution of density and temperature from EBTEL agrees to a reasonable degree with spatially averaged values from the gas dynamic solutions.  The latter are far more costly and would be difficult to apply to many thousand distinct loops.  The EBTEL, on the other hand, requires integration of equations (\ref{EBTEL_eq1}) and (\ref{EBTEL_eq2}), which can be done quickly for each loop in a thousand-loop flare.  The only inputs required
by the model are the loop half-length $L$ and time-dependent heating function $Q(t)$.  Each solution begins with the loop at an initial density, $n_{0}$ and temperature, $T_{0}$, corresponding to a steady pre-flare background heating 
$Q_{bk}$.

\subsection{Constructing Temporally and Spatially Resolved Heating Functions}

The rapid rise of the UV emission reflects instantaneous chromospheric response 
to coronal energy which reconnection releases into the flux tube as it forms.
Radiative transfer modeling of the lower atmosphere dynamics has shown that the 
timescale of the chromospheric response to energy deposit is within a 
few seconds \citep{Fisher85, Canfield87}. In this flare, the major energy transfer from corona to the lower atmosphere 
during the impulsive phase is by thermal conduction, and the characteristic
timescale of thermal conduction is estimated to be around 10$^{1-2}$ seconds given 
the loop top temperature of $T_a = 5 - 10$~MK, density $n = 10^{8-9}$cm$^{-3}$, and loop half length $L = 50$ Mm. Therefore, the thermal flux reaches the
lower atmosphere nearly instantaneously compared with the UV rise time of a few minutes. 
On the other hand, energy deposit at the lower atmosphere drives evaporation which in turn 
fills the corona. The coronal sound speed at the temperature of 10$^{6-7}$ K ranges from
100 - 400 km s$^{-1}$, and the measured half length of post-flare loops in the studied 
flare ranges from 50 - 100~Mm. These values yield the acoustic transit time of order
10$^{0-1}$ minutes. It typically takes a few acoustic transit times to equilibrate
between the corona and transition region, whereas the rise time of the UV emission is within 
10 minutes (see Figure~\ref{Fig2}). These three different timescales, the timescales of 
energy deposit (by thermal conduction) and lower-atmosphere response at a few seconds,
the UV rise time of $<$10 minutes, and the corona plasma hydrodynamic evolution timescale of $\ge$10 minutes,
are sufficiently separated to make it reasonable to assume that the rapid rise of the UV emission 
characterizes the time profile of the heating rate alone.

We therefore use the parameters from fits to the UV rise phase of pixel $i$, denoted $I_i$, 
$t_i$ and $\tau_i$, to define the heating function for the overlying flux tube,
$$H_i(t) = \lambda [I_i q(|t - t_i|/\tau_i)]^{\alpha} {\rm ergs~ s^{-1}~ pxl^{-1}}  ~~.$$ 
 The normalized shape function used to fit the UV rise profile, generically denoted $q(t)$, is also used in the heating profile. In this study, we experiment with the same two functions, a half gaussian and a linear function, used illustratively in the fitting in Figure ~\ref{Fig2}.  
We assume, however, that the heating profile is symmetric; $q(t)$ is a full Gaussian 
 $$q(t) = {\rm exp} [\frac{-(t-t_i)^2}{2\tau_i^2}],  (0 < t < \infty)~~,$$
or a full triangle
$$q(t) = 1 - \frac{|t - t_i|}{\tau_i},  (-\tau_i < t - t_i  < \tau_i)~~,$$
even though the fit used only the rising half of $q(t)$.
The scaling factor $\lambda$ and the power-index $\alpha$ are empirical constant parameters 
by which we will relate the UV count rate to the actual heating rate. In the present study, 
$I_i$, $t_i$, $\tau_i$ are determined directly from observations, and $\lambda$, $\alpha$ 
and the EBTEL parameter $c_1$ are set by comparing observed coronal radiation with the 
results of the EBTEL model. In future work, we intend to determine 
$\lambda$ and $\alpha$ from physical models describing how the UV radiation relates to the heating rate. 

The total heating rate of the flare is given by $H(t) = \Sigma_i H_i(t)$ (ergs s$^{-1}$).
The time integral of the heating rate $W_{in} = \int H(t) dt$, is the total input energy converted to heat by reconnection.  This net energy release can be computed from the observed radiation from the flare. This general 
energy conservation principle guides the initial choice of $\lambda$ when constructing the heating function. 

EBTEL's ad-hoc volumetric heating rate is set to $Q_i(t) = H_i(t)/L_i + Q_{bk}$, where $L_i$ is the half length of the $i$th loop and $H_i(t)$ is the heating function described above. For this flare, we estimate the loop length from the average foot-point separation seen in UV images.
It is seen that from 7:30 to 8UT, the mean separation $D$ between newly brightened pixels in the positive magnetic field and those in the negative magnetic field increases from 40\arcsec\ to 60\arcsec, and from 8:00 to 10:00 UT,  the mean separation $D$ grows from 100\arcsec\ to 160\arcsec. These correspond to two sets of coronal loops
seen in EUV 171\AA\ images in Figure~\ref{Fig1}. Estimates of the mean
distance between the feet of post-flare EUV loops are consistent with the values we derived
using UV observations that look at the foot-points. We assume that the loop is a 
semi-circle so that the half length is given by $L_i = \pi D_i/4$. We also assume that, for each
of the two sets of the loops, the post-flare loop length (or the foot-point separation)
grows linearly with the time of formation. Namely, flare loops formed later result from magnetic reconnection 
occurring at higher altitudes and are therefore longer. This is consistent with observations
of numerous flares including this flare. With these, the loop half length is related to when reconnection takes place,
and ranges from 30 to 100 Mm. $Q_{bk}$ is a very low constant background heating rate, typically taken as
$Q_{bk} = 10^{-4}$ ergs s$^{-1}$ cm$^{-3}$, which is about one thousandth of $H_i/L_i$. 

\subsection{Evolution of Flare Plasmas}
\begin{figure}
\epsscale{1.0}
\plotone{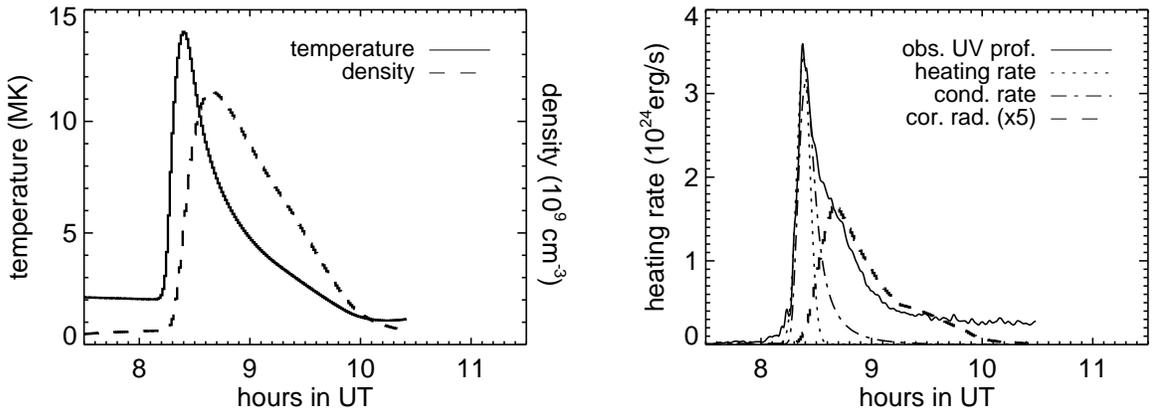}
\caption{Computed evolution of plasma properties in a flare loop rooted
at one flaring pixel. Left: computed time profiles of temperature (solid) and density (dashed) of a flare loop anchored to
a flaring pixel. Right: observed UV 1600\AA\ count rate light curve (solid), the constructed heating function (dotted), and
computed coronal radiation rate (dashed) and conduction rate (dot-dashed) in the flare loop
of cross-sectional area 1\arcsec\ by 1\arcsec.}\label{Fig3}
\end{figure}

From the inputs described above we compute the coronal response solving the EBTEL equations, (\ref{EBTEL_eq1}) and (\ref{EBTEL_eq2}), for each UV-brightened pixel.  Figure~\ref{Fig3} shows the results of the process for a single pixel to which is rooted a single flux tube
(note that here we use ``flux tube" to refer to the coronal structure rooted at a single
pixel of 1\arcsec\ by 1\arcsec, though in reality, a flux tube may be smaller or larger than this size).
The right panel shows the UV light-curve (solid) and heating rate $H_i(t)$ (dotted) derived from it using a Gaussian profile and parameters $\alpha = 1$, $\lambda = 2.7 \times 10^5$ erg/DN.  The panel on the left shows the temperature and density computed from EBTEL with $c_1 = 1.4$.  The impulsive energy input results in plasma evolution on much longer time scales.   Finally, the thermal conduction and coronal radiation from the EBTEL model are plotted against the input curves in the right panel.   
During the impulsive heating, the major energy transport mechanism is by conduction from the corona to the lower atmosphere, and the time profiles of
temperature and conduction flux are nearly identical to that of the heating function. Consequently the lower
atmosphere radiation as reflected in the UV emission also follows the heating function, which in turn
justifies our practice of using the rise phase of the UV light curve to infer the heating function.
On the other hand, the density rises more gradually by chromospheric evaporation. There is a long decay 
in the coronal radiation response which resembles the decay phase of the UV light curve, in agreement with the 
pressure gauge picture of \citet{Hawley92}.  Effects on the flux tube evolution of varying the model's free parameters, $\lambda$, $\alpha$ and $c_1$  will be discussed in the next section.

\begin{figure}
\epsscale{1.0}
\plotone{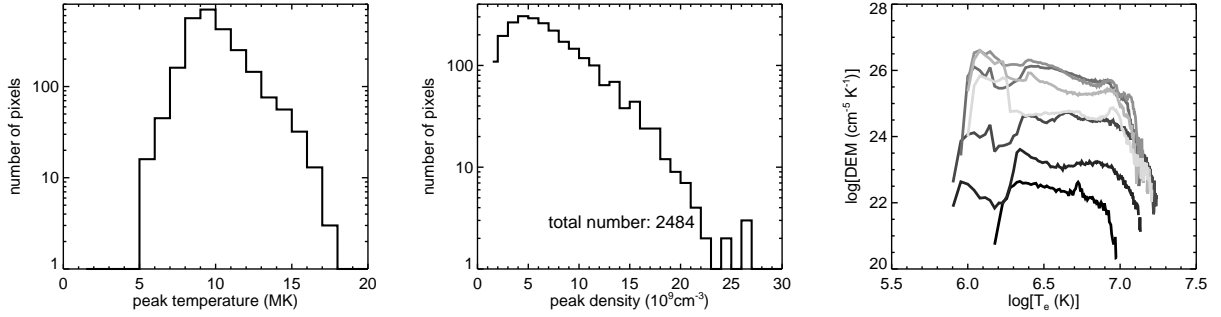}
\caption{Left: peak temperature distribution of 2484 flare loops rooted in 2484 pixels (1\arcsec by 1\arcsec).
Middle: peak density distribution of 2484 flare loops rooted in 2484 pixels. 
Right: the coronal DEM averaged over every half an hour from 7:00 to 10:30 UT (dark to grey).}\label{Fig4}
\end{figure}

The computation illustrated above is repeated for about 2500 flux tubes /pixels identified and characterized by UV observations. These tubes are formed and heated at different times during the flare and by different amount of energy. Figure~\ref{Fig4} shows the distribution of the peak temperature and density of these flux tubes. For this flare, the peak temperature of most flux
tubes ranges from 7 to 15 million Kelvin degrees, and the peak density ranges from 3 - 20
$\times$ 10$^{9}$ cm$^{-3}$. Note that flare loops achieve peak temperature and density at different times depending on when the heating starts and by how much the loop is heated, 
as inferred from $t_i, \tau_i$ and $I_i$. At any given time, we can also compute the differential
emission measure (DEM) of the coronal plasma, as primarily determined by $T_e$ and $n_e$ of numerous loops at different evolutionary stages. The right panel in Figure~\ref{Fig4} shows the
DEM of coronal plasmas averaged every 30 minutes during different phases of the flare.
It is seen that during the rise of GOES emission, the DEM at high temperature
is significant, while during the decay phases, the DEM steepens with a large lower temperature
component, which reflects cooling of a large amount of plasmas.

\subsection{Comparison between Model and Observations}
\begin{figure}
\epsscale{1.0}
\plotone{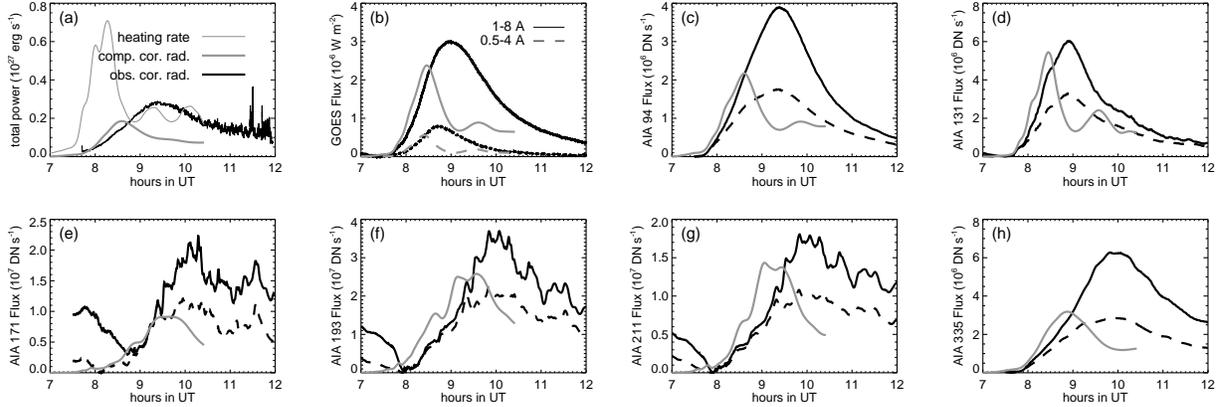}
\caption{Comparison of computed and observed soft X-ray and EUV fluxes. (a) constructed total heating rate (light grey), computed total coronal radiation
(dark grey), and coronal radiation calculated from GOES observation (dark) of the 2010 August 1 C3.2 flare. 
(b): comparison of the soft X-ray fluxes at 0.5-4~\AA\ (dotted) and 1-8~\AA\ (solid) that are computed 
using EBTEL model (grey) and actually observed by GOES (dark).
(c-h): comparison of EUV fluxes that are computed (grey) and observed by SDO/AIA (dark) in six bands.
The dark solid line shows the sum of the observed EUV fluxes in two flare regions indicated in Figure~\ref{Fig1}, and the dark
dashed line shows the observed EUV flux in the top region.}\label{Fig5}
\end{figure}

The coronal DEM described above can be convolved with instrument response functions to compute 
the predicted soft X-ray and EUV fluxes for direct comparison to GOES and AIA observations. 
For EUV flux calculation, we have used the lately updated AIA response functions that are calibrated against EVE and with
the updated CHIANTI database \citep{Boerner2012}. The results of this convolution are shown in Figure~\ref{Fig5}. 
The GOES fluxes are presented with the pre-flare background subtracted, and EUV fluxes are computed
by summing up counts in two flare regions denoted by the two boxes in the lower left panel of
Figure~\ref{Fig1}. In each region, the pre-flare counts are subtracted.

The comparison between model and observation guides our selection of the best-fit parameters.
The model results shown in Figure~\ref{Fig5} are obtained with a gaussian heating function plus a constant 
background heating, and with $\alpha = 1$, $\lambda = 2.7 \times 10^5 $ erg/DN, and $c_1 = 1.4$. 
Specifically, the scaling parameter $\lambda$ has been adjusted to best match the GOES 1-8~\AA\ flux, 
and the parameter $c_1$ is adjusted to match the 171~\AA\ flux. With this choice of parameters we find that the 
peak heating rate of the flux tubes, assumed to be directly proportional to the peak UV counts, 
ranges from 0.02 - 0.2 ergs s$^{-1}$ cm$^{-3}$. The best-fit $c_1 = 1.4$ also suggests that, 
in this small C3.2 flare which is primarily a thermal flare with direct heating taking place in 
the corona, the mean energy loss rate through the transition region is comparable with the coronal radiation loss. 
Comparison of the model results with different sets of parameters will be discussed in the next section.

Our experiments yield computed fluxes in reasonable agreement with the observed rise phase 
of the flare in multiple soft X-ray and EUV bands, though the model parameters are adjusted 
using only observations from two wavelengths. Our adjustments are made using GOES observations, usually 
given off by flare plasmas at relatively high temperatures from a few to 20 million K degrees,
and the AIA 171\AA\ channel, from plasmas of 1-2 MK.  Unbiased comparison may then be made to
fluxes in other AIA EUV bands from intermediate temperatures of a few million K degrees. 
The reasonable agreement in all these wavelengths suggests that our approach 
is capable of capturing heating and evolution during the rise phase of the flare.
Note that in three EUV channels, 171\AA, 193\AA, and 211\AA, the flux decrease in the pre-flare
phase is caused by eruption of pre-existing loops (or dimming), which is then
followed by formation of new post-flare loops seen in the 171\AA\ images in Figure~\ref{Fig1} as well
as reproduced by the model (see the enhanced emission of the new loops at around 9 UT in 171 \AA).

We note that the flare is composed of two distinct loop systems, a set of shorter loops
in the north and a set of longer loops in the south. The two boxes in the lower-left
panel in Figure~\ref{Fig1} nearly include these two sets. In Figure~\ref{Fig5}, we also
examine the observed contribution to the total EUV flux from these two sets of loops. It is seen
that shorter loops evolve more quickly and have dominant contribution to EUV fluxes in the early
stage. Roughly speaking, at relatively high temperatures featured in 94\AA, 131\AA, and 335\AA\ 
bands, emission from the northern region (mostly short loops) dominates until 8:30 UT, and at relatively low temperatures
in other EUV bands (171\AA, 193\AA, and 211\AA), emission from the northern region dominates 
until 9:30 UT. During this early stage (up to 8:30 UT for high temperature emissions and to 9:30 UT
for low temperature emissions), the computed fluxes in nearly all EUV channels match the sum of both sets 
of loops. During the peak and decay of each EUV band, the observed fluxes from the northern and southern
loops are comparable, and the comparison shown in Figure~\ref{Fig5} suggests that the model under-estimates EUV fluxes
from both the northern and southern regions. The experiments in the next sections suggest
that the observed abundant X-ray and EUV emissions during the peak and decay are most 
likely produced by additional heating events not captured in our present method of building heating functions
from the UV 1600\AA\ signatures.

\section{Effects of Varying Model Parameters}
Our method does require the adjustment of a few free parameters. 
The $\alpha$ parameter characterizes how the amount (or magnitude) of UV emission is scaled 
to the energy release rate. In principle, this parameter depends on the detailed physics of the mechanism of lower
atmosphere heating and radiative transfer. $\lambda$ is a scaling parameter that converts the observed, instrumental
data counts to energy in units of ergs. It depends on not only the mechanism of
UV radiation upon atmosphere heating, but also the instrument response function. 
The parameter $c_1$ is a dimensionless constant characterizing the ratio of the total energy loss 
through the transition region to the radiation loss in the corona. This ratio depends on complex coupling of
corona and transition region physics. It is computed in some 1D hydrodynamic coronal heating models
to range from 2 to 20 and vary during different stages of coronal plasma evolution \citep{Klimchuk08}. 
In our experiment, though, we take this parameter as a constant indicative of the mean energy loss through 
the transition region relative to the coronal radiative loss.

\begin{figure}
\epsscale{0.7}
\plotone{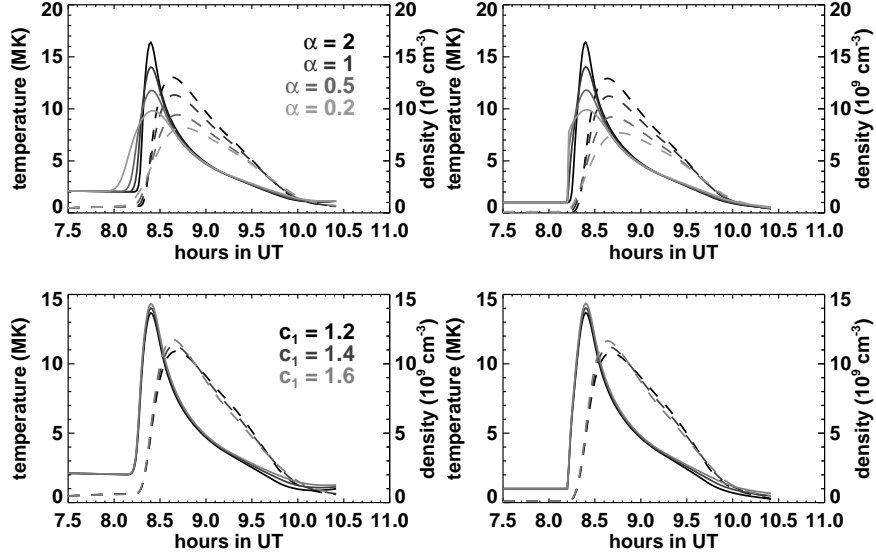}
\caption{Comparison of plasma temperature (solid) and density (dashed) evolution in a flare loop with different 
model parameters. In the left panels, the heating function is a Gaussain with background heating, and in 
the right panels, the heating function is a gaussian without background heating. The top two panels show the
cases with different shape of the heating function ($\alpha = 2, 1, 0.5, 0.2$). For different $\alpha$, the 
best-match $\lambda$ varies, and constant $c_1$ = 1.4 is used. The bottom two panels show plasma properties 
with different $c_1$ value ($c_1 = 1.2, 1.4, 1.6$) that scales the transition region loss to the coronal 
radiation, and with the same $\lambda = 2.7 \times 10^5$ erg/DN and $\alpha = 1$.}\label{Fig6}
\end{figure}

\begin{figure}
\epsscale{0.7}
\plotone{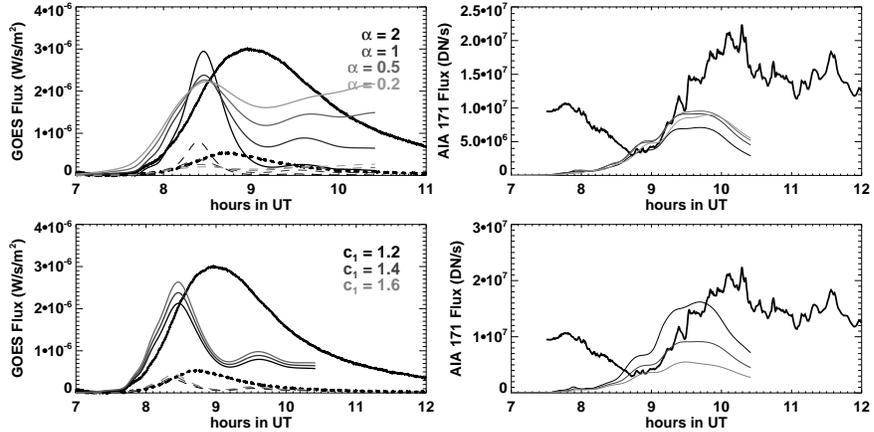}
\caption{Comparison of computed and observed soft X-ray and EUV fluxes with different model parameters. 
The heating function in these panels is a gaussian with background heating. 
The top two panels show the soft X-ray and EUV fluxes for different shapes of the 
heating function ($\alpha = 2, 1, 0.5, 0.2$) with varying $\lambda$, and the bottom two panels 
show the cases for different $c_1 = 1.2, 1.4, 1.6$ values with $\lambda = 2.7 \times 10^5$ erg/DN and $\alpha = 1$.}\label{Fig7}
\end{figure}

\begin{figure}
\epsscale{0.70}
\plotone{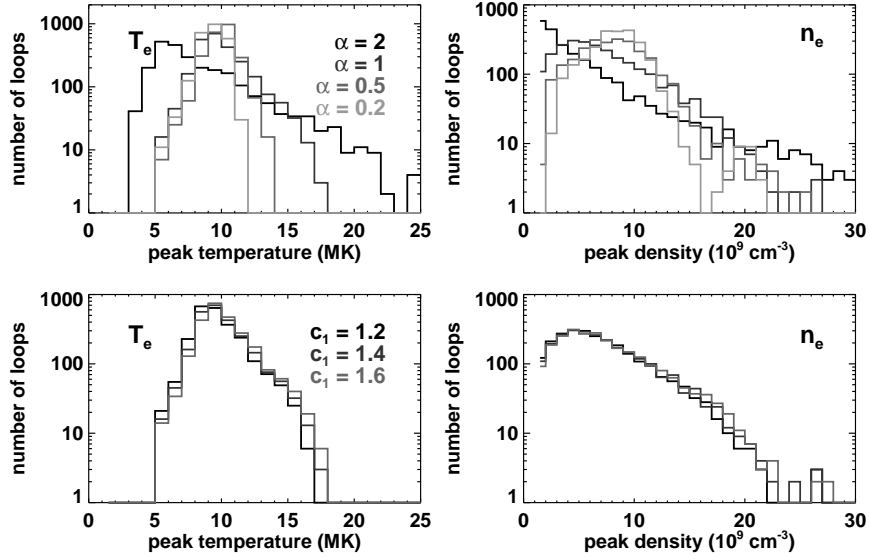}
\caption{Comparison of the peak temperature and density distribution of 2484 flare loops with different model
parameters, with a heating function being a gaussian with background heating.
The top two panels show the cases with different $\alpha = 2, 1, 0.5, 0.2$, and the bottom two panels
show the cases with different $c_1 = 1.2, 1.4, 1.6$.}\label{Fig8}
\end{figure}

Absent a physical model to determine these parameters, it is important to 
understand how sensitive the computed coronal radiation is to the different choices of 
these parameters. In Figure~\ref{Fig6}, we demonstrate how different model parameters 
affect plasma evolution (in terms of temperature and density)
in a single flux tube. In Figure~\ref{Fig7}, we examine how the synthetic coronal radiation
is changed with different model parameters. For simplicity, we only present
the model-observation comparison in GOES soft X-ray channels (high temperature)
and AIA 171\AA\ channel (low temperature). In Figure~\ref{Fig8}, distribution of plasma
properties will be presented.

First, we vary the initial temperature $T_e = 0.5, 1, 2$ MK, and initial 
density $n_e = 0.01, 0.1, 1 \times 10^{9}$cm $^{-3}$ in the flux tube. It is found that 
these initial parameters have no impact on the flux tube evolution as soon as impulsive heating occurs. 
This is a notable contrast to the nanoflare cases explored by \citet{Klimchuk08} in which initial conditions had a significant effect.  The difference is the much stronger heating rate during the flare that overwhelms the initial state leaving no visible difference in the coronal evolution.

We secondly consider the effect of the shape of the heating function in terms of whether 
the heating function is a gaussian or a triangle, whether there is constant background heating
in addition to the impulsive heating, and the variation of $\alpha$ value.  
It is found that whether the heating function is a gaussian or triangle makes some difference 
in the evolution of a single flux tube, and obviously how big the difference is depends on how 
close the gaussian and triangle functions are to each other. However, such difference in single flux tubes
is insignificant when we compare the total radiation by all flux tubes.

Comparing cases with and without background heating (Figure~\ref{Fig6}), 
it is found that, before the impulsive heating, the low constant background 
heating would, regardless of the initial temperature and density, bring the flux tube to a new pre-flare
equilibrium at $T_e \approx 2$~MK, and $n_e \approx 10^{8}$cm$^{-3}$, 
which the flux tube stays with until impulsive heating sets in. 
Once the impulsive heating starts, there is no difference in the evolution of the flux tube as it
is dominated by the strong heating. The difference will show up only during the decay phase
long after the impulsive heating has finished. The continuous constant background heating
will slightly raise the temperature of the flux tube later in the life of the flux tube 
compared with the case of no persistent background heating. As a result, the total coronal 
radiation in the decay phase is more smooth. 

It might appear desirable to raise the background 
heating so as to reproduce the decay phase of the flare which is observed to last longer than modeled; 
however, increased background heating would
significantly reduce the EUV radiation observed in 171\AA\ band, which is sensitive to low
temperature plasmas at around 1-2~MK. Therefore, the experiment suggests that the observed abundant
fluxes in the late phase of the flare may not be produced by continuous heating in the same
flux tubes but by some other yet unknown mechanisms, such as formation of and energy release in new
flux tubes though not detected in the foot-point UV radiation.

We also vary the empirical law relating heating rate with the UV count rate 
by varying the power index $\alpha = 2, 1, 0.5, 0.2$. The effects are 
obvious and predictable in several respects. With greater $\alpha$, the heating is more impulsive and the 
resultant flux tube evolution is more rapid. Because of
different $\alpha$ used, we also have to modify the scaling factor $\lambda$ for the best match. 
The top two panels in Figure~\ref{Fig6} and in Figure~\ref{Fig8} show that with greater $\alpha$, higher peak temperature and peak density are achieved, but the timings of the peak temperature and density do not change.
It is also remarkable that the temperature and density attained through these quite different heating functions
all converge in the late decay phase.

When the total coronal radiation flux is computed, it is seen in the top panels in Figure~\ref{Fig7} 
that the GOES radiation fluxes evolve on shorter time scales for larger value of $\alpha$, 
but the peaking times remain the same. This result suggests that the timing of the coronal radiation
critically depends on the temporal distribution of the heating functions, namely, when flare loops
are formed and heated, but not the exact shape of the heating functions. The top right panel in Figure~\ref{Fig7}
shows the comparison between observed and computed EUV flux at 171\AA.  It is noteworthy that
the difference produced by different $\alpha$ values is much less significant than in the soft X-ray
fluxes. Specifically, the duration of the EUV flux at low temperature is not sensitive to $\alpha$ value at all.

Both effects are consistent with the fact that the flux tube evolves to the same temperature and density
in the late decay phase regardless of the shape of the heating function.
This result suggests that the exact shape of the heating function has some impact on the duration 
of high-temperature plasma radiation, when conductive cooling time scale matters, but has little impact on the duration of radiation
by low-temperature plasmas which have cooled down, been ``smoothed" out, and in a sense lost the 
memory of the heating function. 

Last, we alter the parameter $c_1$ which in the EBTEL model scales the transition region loss
to the coronal radiation. Since the net input energy should equate the net output energy,
which is the sum of coronal radiation and transition region loss, a greater $c_1$ value requires
a larger heating rate. For this C3.2 flare, the optimal value of $c_1$ is 1.4, much smaller
than the quoted $c_1 = 4$ in \citet{Klimchuk08}. Our experiments suggest that the computed 171\AA\ flux
is very sensitive to the choice of $c_1$, a slightly greater value of $c_1$ by 15\% would bring down
the computed 171\AA\ radiation flux by a factor of 2. This is shown in the bottom right panel of Figure~\ref{Fig7}.

From the bottom panels in Figure~\ref{Fig6}, it is seen that a different $c_1$ does not modify the temperature and
density except during the very late phase, when a larger value of $c_1$ leads to higher temperature
of the flux tube. During this phase, radiative cooling dominates in the corona, whereas
the coronal plasma density is decreasing by draining (i.e., the downward enthalpy flow).
A greater transition region loss speeds up the coronal draining so that coronal
density is lower, and in turn slows down the radiative cooling. This effect accounts
for the higher coronal plasma temperature with larger $c_1$ value.
It is evident that 171\AA\ radiation flux is by plasmas cooled
to nearly 1MK, and the higher plasma temperature in the late decay 
phase would significantly decrease the radiation flux in this wavelength.

\begin{figure}
\epsscale{0.8}
\plotone{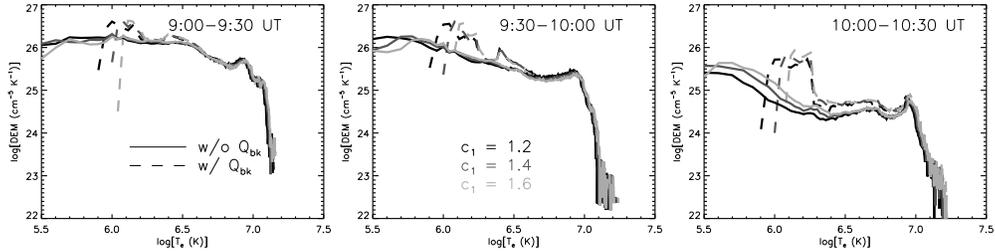}
\caption{Comparison of the time-averaged coronal DEM computed with different $c_1 = 1.2, 1.4, 1.6$
and with or without constant background heating
in three periods, 9:00 - 9:30 UT, 9:30 - 10:00 UT, and 10:00 - 10:30 UT, when the EUV 171\AA\ radiation is prominent. }\label{Fig9}
\end{figure}

To further examine this, we specifically sample the late phase of the flare, and compute the DEM averaged during 9:30 - 10:00 UT
and during 10:00 - 10:30 UT, when the observed 171 \AA\ radiation flux is prominent. Figure~\ref{Fig9}
shows that for different $c_1$ values, the DEM at 1-2 MK is very different. Specifically, for greater
$c_1$, the DEM at 1-2~MK is smaller. This is the temperature that primarily contributes to 
radiation at 171\AA\ band. 

There is no visible difference at higher temperatures; therefore, 
the soft X-ray radiation flux emitted by plasmas at higher temperatures (such as the flux
observed by GOES) does not vary much with varying $c_1$ values. In the figure, we also
contrast cases with and without background heating, which, not surprisingly, has
a strong impact on the low temperature DEM but little impact on the high temperature DEM.

In summary, the analyses above suggest that model parameters affect modeled radiation at
different temperatures in different ways. Among the few free parameters, the
shape of the heating function has more impact on the evolution timescale of
high-temperature plasma radiation, whereas low-temperature plasma radiation
during the decay phase is more sensitive to the background heating and the transition
region loss rate. Therefore, utilizing soft X-ray and EUV observations
at multiple wavelengths will eventually enable us to fully constrain the few
free parameters ($\lambda$, $\alpha$, and $c_1$) describing the heating
functions, in addition to the other parameters ($t_i$, $\tau_i$, $I_i$, $L_i$)
that are directly measured from observations.

\section{CONCLUSIONS AND DISCUSSIONS}

\subsection{Summary of Results}
In this paper we present a method to infer from observations the temporal distribution of
heating rates in numerous flux tubes (flare loops) formed by reconnection
and subsequently heated. We assume that the rapid rise of the UV emission from the lower atmosphere
is the instantaneous response to heating in the flare loops. Therefore,
the rise phase of the spatially resolved UV light curve is used to construct
heating rates in individual flare loops, which describe when, for
how long, and by how much flare loops are heated. 
The observationally inferred heating rates are used in
the EBTEL model to compute evolution of coronal plasmas and the total radiation
flux by all these tubes at different evolution stages at any given time.
The method is applied to a C3.2 flare observed by SDO on 2010 August 1, and
the computed soft X-ray and EUV fluxes agree well with those observed by GOES and AIA
during the rise phase of this long-duration flare. The preliminary results suggest 
that this method may have captured the distribution of impulsive heating rates during
the rise of the flare.

We also examine how model results change with varying model parameters to gain
insight concerning what physical process is critical in different stages
of flare plasma evolution. The experiment suggests that, within the current
frame work, computed radiation at different temperatures is sensitive to different
model parameters. The existing instrument capabilities are able
to obtain observations of flare plasmas at many different temperatures. Utilizing
these will certainly provide more observational constraints than free model parameters,
and therefore help determine some of the physical quantities used in the model.
For example, the present study may be extended to computing predicted (high temperature)
plasma radiation spectrum and light curves observed by RHESSI \citep{Liu2012}, 
as well as computing the intermediate temperature radiation spectrum by EVE. 

Our experiments also show that the peak time of coronal radiation at different
temperatures is primarily determined by the distribution of heating events which can be
constrained by observations, but barely change with other free parameters used in the model. Therefore, the missing radiation during
the peak and decay phase may indicate additional heating events during the late phase of the flare
that are not detected with our current approach of analyzing UV foot-point radiation signatures. 
In the next Section, we will discuss on these missing heating events.

\subsection{Missing Heating Events in the Late Phase}
To find out the additional heating events possibly missed by our method, we 
re-examine the UV light curves in individual pixels. Our present approach to reconstruct 
heating functions employs only the rise of the UV emission by considering that the gradual smooth
decay is governed by the coronal evolution. Closer scrutiny reveals that some of the flaring
pixels exhibit a second emission peak after the first peak. However, accurately characterizing the second
peak in the decay phase is difficult, since it is mixed with the decay of the first
peak. For a first-order examination, we subtract the first peak, which is the full-gaussian heating 
function from the fit, from the UV light curve, and then search for the second peak 
that occurs more than 2$\tau$ after the first peak, $\tau$ being the Gaussian width of the the first peak. 

\begin{figure}
\epsscale{1.0}
\plotone{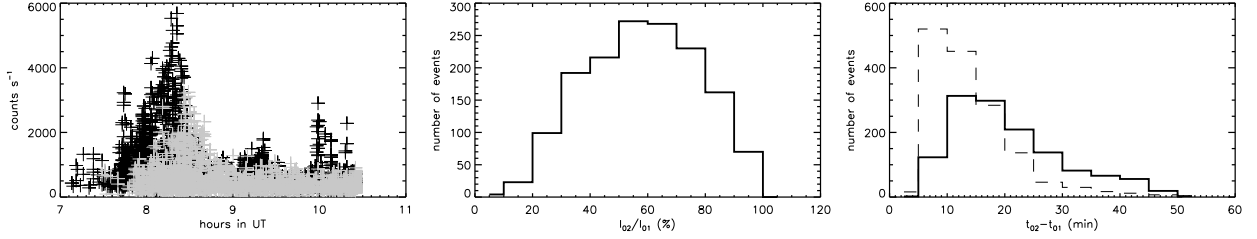}
\caption{Left: the peak counts rate and peak time of the first (dark) and second (grey; see text) components
in the UV light curves of the flaring pixels. Middle: histogram of the ratio of the peak counts rate of the second component to that of the
first component. Right: histogram of the time lag of the peak time of the second component relative
to the peak time of the first component. Superimposed in dashed line is the histogram of twice the rise time
of the first UV component.}\label{Fig10}
\end{figure}

Figure~\ref{Fig10} shows the statistical properties of so-derived secondary peaks.
The left panel shows the peak counts rate and time of the first (dark) and second (grey)
peaks. The middle panel presents the histogram of the ratio of the peak counts
rate of the second peak to that of the first peak, with the median value at 60\%. The right panel
shows the histogram of the time lag of the secondary peak relative to the first one (solid), 
compared with 2$\tau$ (dashed) of the first peak. From this plot, the median value of time lag
is 20 minutes; in comparison, the median of $2\tau$ is 12 minutes. 
This simple experiment also reveals that the so-derived secondary peak may be present in about 60\% flaring pixels.
We must note that, on the one hand, some of these secondary peaks may still represent the decay of the
first heating rather than real heating events, especially if they occur very close to the time of the first peak; on the other hand,
if additional heating exists but is rather gradual with a low peak magnitude, it cannot be 
recovered from this experiment. 

Lacking an approach to accurately pick out all additional heating events and characterize their properties, such
as the heating duration, we assume that the secondary heating in general takes place at 20 minutes after 
the first heating and with the peak counts rate about 60\% of the first peak counts rate. These values 
are median values from the histograms shown in Figure~\ref{Fig10}. We further assume that the duration of these
events is the same as the first heating. Also, we assume that the secondary heating takes place not in the same
flux tube that is heated by the first heating event, but in the adjacent flux tube which is observationally
unresolved from the primary flux tube. Such is necessary because continuous heating of the same flux tube 
would not allow the flux tube to cool down to emit at low temperatures such as in AIA 171\AA, as revealed 
by our fore-going parameter runs. Therefore, the secondary peaks are treated as new heating events in 
new flux tubes, whose evolution can be computed by EBTEL model.

\begin{figure}
\epsscale{1.0}
\plotone{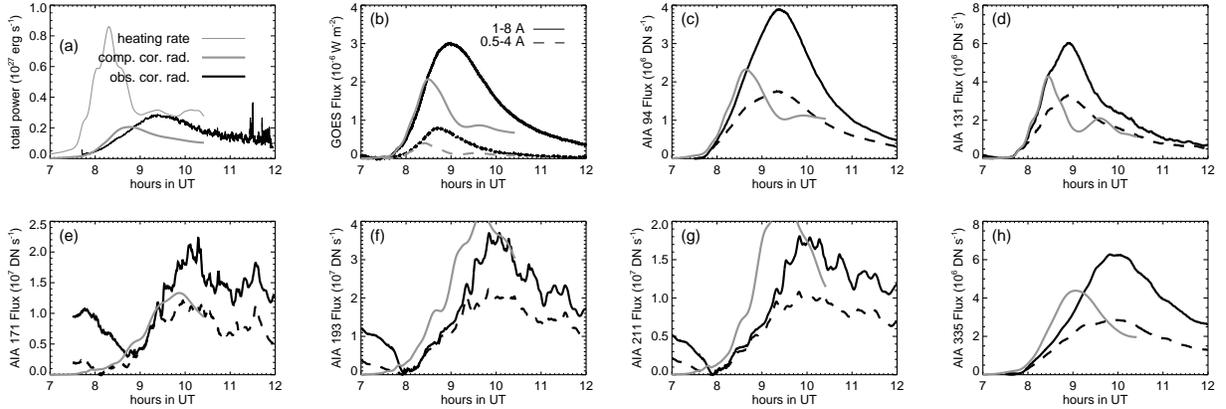}
\caption{Comparison of computed and observed soft X-ray and EUV fluxes taking into
account artificial secondary heating components and heating of half loops (see text). 
These are computed using $\alpha = 1$, $\lambda = 2.2\times 10^5$ erg/DN and $c_1 = 1.4$.
(a) constructed total heating rate (light grey), computed total coronal radiation
(dark grey), and coronal radiation from GOES observation (dark) of the 2010 August 1 C3.2 flare. 
(b): comparison of the soft X-ray fluxes at 0.5-4~\AA\ (dotted) and 1-8~\AA\ (solid) that are computed 
using EBTEL model (grey) and actually observed by GOES (dark). (c-h): comparison of EUV fluxes that 
are computed (grey) and observed by SDO/AIA (dark) in six bands. The dark solid line shows the sum of
the observed EUV fluxes in two flare regions indicated in Figure~\ref{Fig1}, and the dark
dashed line shows the observed EUV flux in the top region. }\label{Fig11}
\end{figure}

Figure~\ref{Fig11} shows the comparison between observed and synthetic soft X-ray and EUV fluxes with
the increased number of heating events. Compared with Figure~\ref{Fig5}, the additional heating events
certainly produce more radiation leading to better agreement with the observed fluxes, especially
the EUV fluxes observed by AIA - we recognize that the abundant late-phase EUV fluxes in
some AIA channels, notably 94\AA, 131\AA, and 335\AA, may be emitted by very low temperature plasmas of $<$ 1MK, which cannot
be reproduced by the EBTEL. However, these additional heating events do not appear sufficient to bring 
up the amount of high temperature soft X-ray flux during its peak and decay.

\begin{figure}
\epsscale{1.0}
\plotone{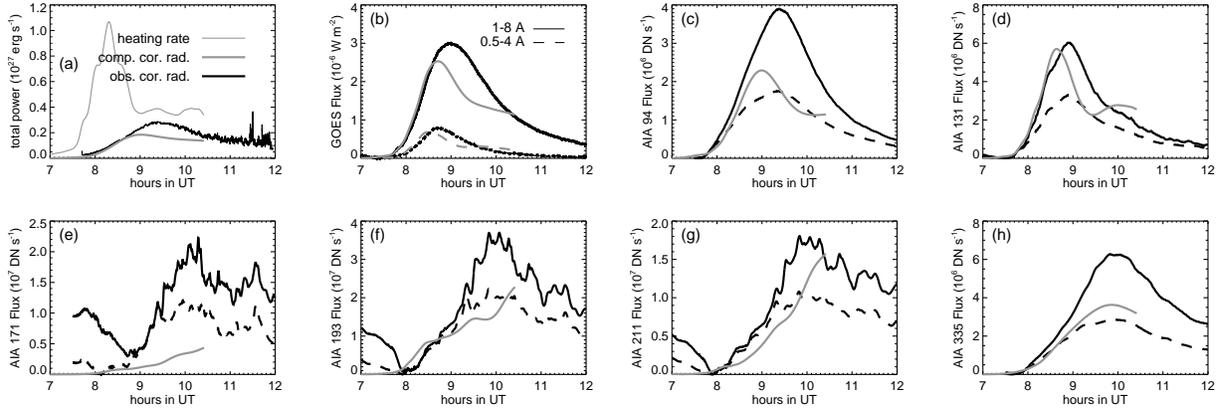}
\caption{Same as Figure~\ref{Fig11}, but for the case of heating full loops (see text). These are
computed using $\alpha = 1$, $\lambda = 2.8\times 10^5$ erg/DN and $c_1 = 1.4$.}\label{Fig12}
\end{figure}

As an alternative, we consider the scenario that flare energy release in the flux tube is 
asymmetric along the tube, such that the downward energy flux primarily deposits at 
only one foot-point. In this scenario, a heating function derived from the foot-point
signature should be applied to heating a full loop rather than a half loop, namely,
the effective loop length is twice as large. In this experiment, we use the EBTEL model 
to compute heating of full loops rather than half loops with the same set of heating functions, 
even though in this case, the underlying assumption of homogeneous heating adopted in the EBTEL 
model is no longer valid. Figure~\ref{Fig12} shows that, using the full-length rather than 
the half-length of the loop certainly changes the evolution timescale, as would be well expected.
Whereas the computed evolution timescale better matches that observed in high temperatures (such as
soft X-ray fluxes by GOES), the predicted evolution timescale for the low-temperature EUV fluxes, 
notably in 171\AA, 193\AA, and 211\AA, appears too long in comparison with observations, or 
plasmas cool down too slowly in these full loops compared with observations.

Comparing the above two experiments shown in Figure~\ref{Fig11} and Figure~\ref{Fig12}, it appears
that still more heating events in the decay phase of the flare would be needed to bring up 
the high-temperature flux to the observed level while also maintaining the low-temperature flux.
Similar conclusion is reached by \citet{Hock2012} in their attempt to reproduce the
second EUV emission peak at $\sim$3MK in some flares, except that in our study, the
new energy release events should make up for high-temperature emissions seen by GOES.

Other possibilities include that the secondary heating events may have longer duration 
than $\tau$ derived from the fit to the first peak. It is also likely that 
the mean-property approach (for example, the constant $c_1$)
of our EBTEL modeling does not accurately describe plasma evolution especially 
during the cooling phase of the flux tube \citep{Klimchuk2012}. To reliably distinguish 
these possibilities requires spatially resolved analysis and modeling beyond the scope
of the present experiments. These experiments, however, provide insight in how plasma evolution would vary
with these different possibilities and the combination of them.

\subsection{The Physical Basis of Heating Functions}
In this paper, the heating function is empirically rather than physically based. 
The model assumes that spatially resolved UV light curves provide time profiles
of energy release rate. The timing information can be justified by the fact 
that the energy transfers from the corona and the lower atmosphere reacts to energy deposit
on very short timescales of only a few seconds. But the amount of UV 
emission depends on complicated physics of lower-atmosphere heating and radiative transfer, so that 
it is usually not simply proportional to the magnitude of energy release rate. In this respect, there 
have been a few studies probing the quantitative relation between UV and hard X-ray emissions. 
\citet{McClymont86} have shown that, for a few tens of flares, the peak EUV (10-1030\AA) flux
is correlated with the peak hard X-ray photon flux by a power-law relation, and applied an 
analytic model to explain the correlation as that EUV emission is produced by particle beams driving
explosive evaporation. Recently \citet{Qiu10} also found a similar power-law relation between the observed
UV and hard X-ray counts during the Bastille-day flare. Such UV-HXR relation provides a reference 
for our experiment, particularly the exercise of varying $\alpha$ values in the heating 
function model. We reckon that such flux-flux relation might not simply 
extend to the pixel-to-pixel variation in the UV emission, or to heating mechanisms other 
than by electron beams. Nevertheless, in this paper, we adopt the simple relation as the first-order 
approach to model a few thousand flux tubes. As the focus of the present study is to 
compare the collective effect by the sum of these few thousand flux tubes, details of pixel-to-pixel
discrepancy are likely unimportant.

It is imperative in future investigations to, first, establish the quantitative relation between UV emission
and energy release rate using physical models computing radiative transfer in the dynamic atmosphere. 
The preliminary results of this kind are shown in \citet{Cheng12}, who have computed the UV continuum 
emission in the lower atmosphere heated by electron beams. Second, distinction between pixels 
(or flux tubes) will become necessary as well as important when computed plasma evolution in 
single loops can be compared with spatially resolved analysis of coronal loops. Such analysis 
is becoming possible with coordinated high resolution observations by AIA at multiple UV and EUV bands. 
Being able to identify individual flare loops from their feet to top and follow their evolution will not 
only help to better determine the heating function of a loop rather than a half loop, 
but also provide insight about the asymmetric energy deposit at the conjugate 
feet of a flare loop, which has been known yet hardly understood for decades \citep[e.g.][]{Fletcher01, Qiu10}.

\acknowledgments We thank Dr. J. Klimchuk for insightful discussion about the EBTEL model,
Dr. R. C. Canfield for illuminating us about the lower-atmosphere radiative transfer, 
and Drs. S. Freeland and G. Slater for help with SDO data acquisition. We thank the referee for very constructive 
comments that help improve the clarity of the manuscript. We acknowledge SDO for providing quality 
observations. This work is supported by NSF grant ATM-0748428 and NASA grant NNX08AE44G.

{}

\end{document}